\documentstyle[twoside,12pt]{article}
\newcommand{\bp} {{\bf p}}

\newcommand{\btab}{\begin{tabbing}}
\newcommand{\etab}{\end{tabbing}}
\newcommand{\eqntimes}{\mbox{} \times}

\newcommand{\beqn}{\begin{equation}}
\newcommand{\eeqn}{\end{equation}}
\newcommand{\barr}[1]{\begin{array}{#1}}
\newcommand{\earr}{\end{array}}
\newcommand{\beqna}{\begin{eqnarray}}
\newcommand{\eeqna}{\end{eqnarray}}
\newcommand{\btablec}{\begin{table} \begin{center}}
\newcommand{\etablec}{\end{center} \end{table}}
\newcommand{\lapprox}{\stackrel{<}{\scriptstyle \sim}}

\newcommand{\gapproxeq}{\lower.7ex\hbox{$\;\stackrel{\textstyle>}{\sim}\;$}}
\newcommand{\lapproxeq}{\lower.7ex\hbox{$\;\stackrel{\textstyle<}{\sim}\;$}} 
\newtheorem{theorem}{Theorem}
\newcommand{\bth}{\begin{theorem}}
\newcommand{\eth}{\end{theorem}}

\newcommand{\plabel}[1]{\label{#1}}
\newcommand{\pbibitem}[1]{\bibitem{#1}}

\newcommand{\ca}{{\cal A}}
\newcommand{\cabc}{{\cal A}_{B \leftrightarrow C}}
\newcommand{\cat}{{\cal A}_{tot}}
\newcommand{\cf}{{\cal F}}
\newcommand{\cfbc}{{\cal F}_{B \leftrightarrow C}}
\newcommand{\cc}{{\cal C}}
\newcommand{\ccbc}{{\cal C}_{B \leftrightarrow C}}
\newcommand{\cz}{{\cal J}} 
\newcommand{\czbc}{{\cal J}_{B \leftrightarrow C}}
\newcommand{\lr}{\leftrightarrow}
\newcommand{\rl}{\leftrightarrow}
\newcommand{\pbx}{\bp \rightarrow -\bp}
\newcommand{\ti}{{\otimes}}

\newcommand{\Eta}{{\Omega}}
\newcommand{\T}{{\Im}}
\newcommand{\figone}{{$1$}}
\newcommand{\figtwo}{{2}}
\newcommand{\figthree}{{3}}
\newcommand{\figfour}{{4}}
\newcommand{\figfive}{{5}}
\newcommand{\figsix}{{6}}
\newcommand{\figseven}{{7}}
\newcommand{\figeight}{{8}}
\newcommand{\fignine}{{9}}
\newcommand{\eh}{{\hspace{.7cm}}}
\newcommand{\ehh}{{\hspace{.3cm}}}
\input epsf

\begin{document}
\title{\begin{flushright}
\small{hep-ph/9608487} \\ \small{MC-TH-96/25} \end{flushright} 
\vspace{0.6cm}  
\LARGE \bf Symmetrization Selection Rules}
\author{Philip R. Page\thanks{E--mail: prp@jlab.org} \thanks{\small \em Present address:
Theory Group, Thomas Jefferson National Accelerator Facility, 
12000 Jefferson Avenue, Newport News, VA 23606, USA.}\\
{\small \em Department of Physics and Astronomy, University of Manchester,} \\
{\small \em Manchester M13 9PL, UK}}
\date{August 1996}
\maketitle
\abstract{We introduce strong and electromagnetic interaction selection
rules for the two--body decay and production of exotic $J^{PC} =
0^{+-}, 1^{-+}, 2^{+-}, 3^{-+} \ldots$ hybrid mesons, four--quark
states and glueballs. The rules arise from symmetrization in states. Examples include various decays to $\eta^{'}\eta,\; \eta\pi,\; \eta^{'}\pi$ and $\pi^{\pm}\pi^{0}$.
The symmetrization rules can discriminate between hybrid and four--quark interpretations of a $1^{-+}$ signal.}

\vspace{1cm}

Selection rules valid for $SU(3)$ flavour symmetry were 
first noted using the Wigner--Eckart theorem \cite{meshkov}, and later
were recognized as being valid for the decay of hybrids  $1^{-+}\rightarrow\eta\pi,\; \eta^{'}\pi$
\cite{peneqcd,lipkin2} within the context of isospin symmetry. 
We offer an approach in which all possible rules of the same kind can be classified. 
$SU(3)$ flavour symmetry will not be assumed, and our selection 
rules do not trivially follow from the reduction of
$SU(3)$ to isospin $SU(2)$ symmetry. The 
$1^{-+}\rightarrow\eta\pi,\; \eta^{'}\pi$ rules follow as a specific example.
We obtain a novel and substantially enlarged list of processes to which selection rules
apply. Both the necessary and sufficient
conditions for the validity of the rules are clearly indicated. 
We also demonstrate that non--trivial rules arise even in the absence of assuming
isospin symmetry. 

In the following we shall be interested in fully relativistic 
two--body strong and electromagnetic decay and production $A\rl
BC$ processes in the rest frame of A. Since strong and electromagnetic
interactions are considered, we assume charge
conjugation $C$ and parity $P$ conservation, but {\it
not} in general isospin symmetry. For simplicity we shall usually
refer to the decay process $A\rightarrow BC$, but the statements will be equally
valid for the production process $A\leftarrow BC$.
We shall restrict the states A, B and C to some assumed leading 
combination of ``valence'' quarks with arbitrary gluonic content, except when
sea components are explicitly considered.
The strong interactions include all interactions
described by QCD. The quarks and antiquarks in A are assumed to travel in all possible 
complicated paths going forward and backward in time and emitting and absorbing
gluons until they emerge in B and C. We shall restrict B and C to angular momentum $J=0$ states with valence $Q\bar{Q}$ quark content and arbitrary gluonic excitation, i.e. to hybrid or conventional mesons. B and C can be radial
excitations or ground states, with $J^{P} =
0^{-}$ or $0^{+}$. If C--parity is a good quantum number, 
$J^{PC} = 0^{-+},0^{+-},0^{++}$ or $0^{--}$ are allowed. Since $0^{-+}$ ground state
meson states B and C are most likely to be allowed by phase space, they are used in
the examples.
We assume that states B and C are identical in all respects
except, in principle, their flavour and their equal but
opposite momenta $\bp_B \equiv \bp$ and $\bp_C \equiv -\bp$.
Hence B and C have the same parity, C--parity, radial and gluonic excitation,
as well as the same internal structure. 

The three symmetrization selection rules for various
topologies are clearly stated in the next section, where we proceed to derive the rules.

\section{Symmetrization selection rules}

For the leading theory of the strong
interactions, QCD, a decay or production amplitude is a linear combination of products
of colour $\cc$ and flavour $\cf$ overlaps, and the ``remaining'' overlap $\cz$.
For reasons that will soon become evident, we shall be interested in
the exchange properties of these overlaps when the labels
(e.g. parity, C--parity, radial and gluonic excitation,
and internal structure) that specify the states B and C are formally exchanged, denoted by $B
\lr C$. For example, $\ccbc$ denotes the effect of exchanging the
colour labels of B and C. 

We are only interested in decays where B and C have the same
colour content, i.e. the way the quarks and gluons 
couple to form the total colour singlet state required by QCD is identical. For a 
conventional meson the quarks and antiquarks 
are in $\bf 3$ and $\bar{\bf 3}$ representations. In an adiabatic 
picture \cite{perantonis90,page95light} the same  holds when B and C are hybrid
mesons. In a constituent gluon picture hybrid mesons B and C have
the colour coupling of an {\bf 8} gluon with $\bf 3$ quarks and $\bar{\bf 3}$ antiquarks. 
As long as the colour content of B and C
is identical we trivially have $\ccbc = \cc$.

When we exchange $\pbx$, we equivalently exchange $\bp_B \lr
\bp_C$. But since all other aspects (other than flavour) of B and C are the same, it is in
fact equivalent to exchanging labels $B  \lr C$ for every
property in the remainder of the state. 
So $\pbx$ is equivalent to $\cz \rl \czbc$.

We shall be interested in processes where the amplitude\footnote{When B and C have
$J=0$, helicity and partial wave amplitudes are identical.} is in principle the
{\it sum of two parts} (or ``diagrams''), i.e. 

\beqn \cat (\bp) = \ca (\bp) + \cabc (\bp) = \cc\ti\cf\ti\cz (\bp)\; + \; \ccbc\ti\cfbc\ti\czbc (\bp)\eeqn
The amplitude is the sum of two parts for the coupling of (hybrid) mesons
and four--quark states shown in Figs. \figone,  \figtwo,  \figfour\ -- \figseven,
since there is always either the possibility that a quark $Q$ in A would end up  
in the particle with momentum $\bp$ {\it and} the possibility that 
it would end up in the particle with momentum $-\bp$, 
corresponding to $ \ca$ and $\cabc$ respectively. 

Under $\pbx$  (or equivalently $\cz \rl \czbc$) 

\beqna \plabel{eqw} \lefteqn{\cat (\bp) \rightarrow 
\cc\ti\cf\ti\cz(-\bp)\; +\ccbc\ti\cfbc\ti\czbc(-\bp)  = 
\cc\ti\cf\ti\czbc (\bp)\; \nonumber } \\ & & +\ccbc\ti\cfbc\ti\cz (\bp)  = 
f\; \{\ccbc\ti\cfbc\ti\czbc (\bp)\; + \cc\ti\cf\ti\cz(\bp)\} = f\cat(\bp)
\eeqna 
where we used $\ccbc = \cc$ and defined $\cfbc \equiv f \cf$. We shall
only be interested in cases where $f=\pm 1$, and where both $\cf$ and $\cfbc$ are non--zero.
If $f = (-1)^{L+1}$, where $L$ is the partial wave
between B and C, it follows from 
Eq. \ref{eqw} that $\pbx$ implies
that $\cat \rightarrow (-1)^{L+1}\cat$. 
 Since in L--wave under $\pbx$ we have by analyticity that $\cat(\bp) \rightarrow (-1)^{L} \cat(\bp)$,
it follows that $\cat(\bp)$ vanishes. This is the symmetrization
selection rule, arizing due to symmetrization in states B and C. 

We  now find necessary and sufficient conditions for the
requirement $f = (-1)^{L+1}$.
Since B and C are identical (except possibly in flavour) they have the same parity, and we conclude that for a parity
allowed process, $P_{A} = (-1)^{L}$. We shall show in 
subsections \ref{ss1} -- \ref{noncon} that for
various flavour scenarios $f = C_{A}^{0}$. For a neutral state,
$C_{A}^{0}$ is just the C--parity of the state. 
For charged states (with no C--parity), we assume that at least one of
the states in the isomultiplet
it belongs to has a well--defined C--parity, denoted by $C_{A}^{0}$. Hence

\beqn P_{A}= (-1)^{L} = - (-1)^{L+1} = -f = - C_{A}^{0}\eeqn
i.e. state A is CP odd.
Since states B and C both have $J=0$, it follows by conservation of
angular momentum that an L--wave
decay would necessitate $J_{A} = L$. Hence states A have $J^{PC} = 
0^{+-}, 1^{-+}, 2^{+-}, 3^{-+}, \ldots$, which are exotic $J^{PC}$
not found in the quark model. So A is not a conventional meson. 

We now show that $f = C_{A}^{0}$.
 
\subsection{Indistinguishable flavours \plabel{ss1}}

The simplest case is when states B and C have {\it indistinguishable flavours}, e.g. $Q\bar{Q}\;  Q\bar{Q}$
states B and C. This does not have to be satisfied for the full
flavour content of B and C, but the decay must be such that
indistinguishable flavour components of B and C are always selected. Then $f=1$. The only interesting cases 
arise when B and C are neutral; and since their C--parities are identical, we have $C_{A} = 1$ by
conservation of C--parity. So $f = 1 = C_{A}$. 
The preceding argument is independent of the decay topology, even though specific examples are
determined by the topology, and are given in the following subsections.

{\bf Symmetrization selection rule I:} (Indistinguishable flavours)
Decay and production in topologies 1 -- 9 (see Fig. 1) to two
$J=0$ hybrid or conventional mesons B and C which are identical in all respects, except possibly flavour,
vanish. This only applies to a $J^{PC} = 
1^{-+}, 3^{-+}, \ldots$ hybrid, four--quark state or glueball A coupling 
to flavour components of B and C that are indistinguishable,
e.g. $Q\bar{Q}$ or $Q\bar{Q}q\bar{q}$
$\rightarrow Q\bar{Q} \; Q\bar{Q}$. Isospin symmetry
is not assumed.

The remaining cases for which $f = C_{A}^{0}$ are discussed in the following
subsections. The corresponding symmetrization selection rules are then stated.

\begin{figure}
\begin{center}
\leavevmode
\hspace{-.3cm}\hbox{\epsfxsize=5 in}
\epsfbox{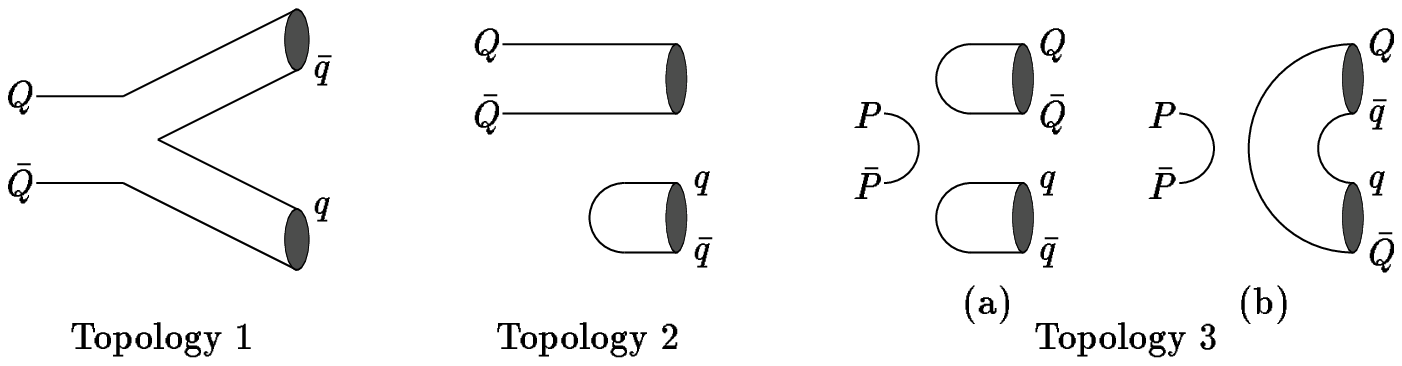}
\vspace{1cm}

\hspace{-.2cm}\hbox{\epsfxsize=5 in}
\epsfbox{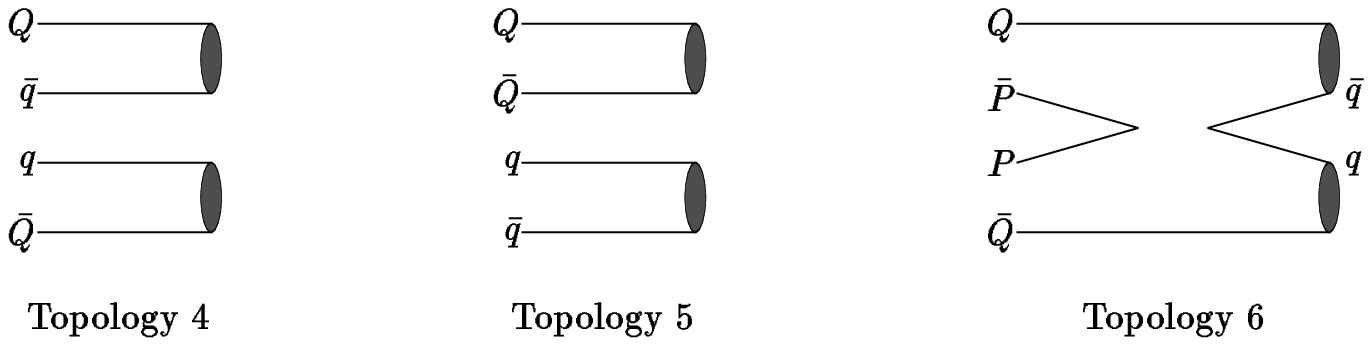}
\vspace{1cm}

\hspace{-.6cm}\hbox{\epsfxsize=5 in}
\epsfbox{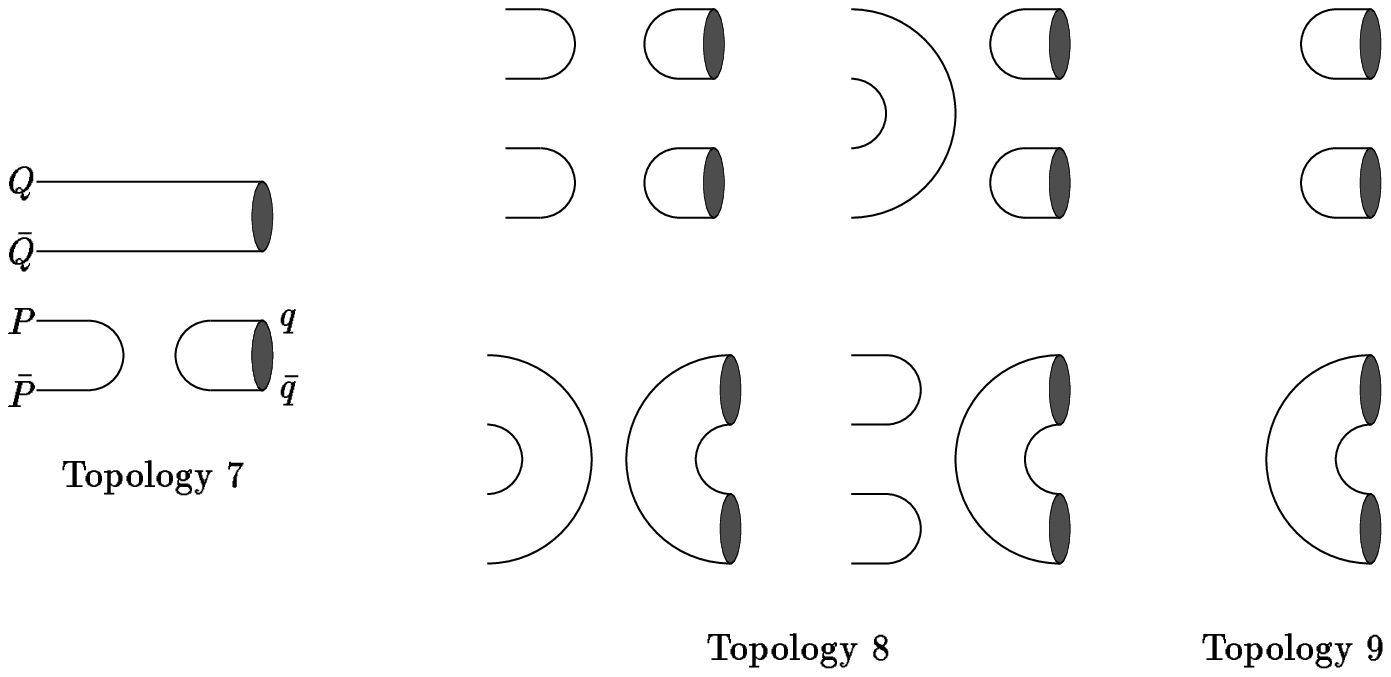}
\vspace{-.6cm}
\caption{Decay topologies. For each diagram state A is on the left--hand side, and
states B
and C on the right--hand side. Quark flavours are
labelled by $Q,\; q$ and $P$.}
\end{center}
\end{figure}

\leavevmode

\subsection{Connected hybrid meson coupling \protect\plabel{hy}}

The possible hybrid decay topologies are
\figone\ -- \figthree, but we focus here on the connected topology \figone.

{\it Indistinguishable flavours:} 
{\sf Examples:} $\Pi^{0} \rightarrow \pi^{0} \eta,\;\pi^{0} \eta^{'}$;\eh  
$\Eta,\; s\bar{s}\rightarrow \eta^{'}\eta$,

\noindent where $\Pi^{0}$ denotes a $u,d$ quark neutral isovector state (e.g. $u\bar{u} - d\bar{d}$), and
$\Eta$ an isoscalar state (e.g. $u\bar{u}+d\bar{d}$). 
In the examples listed the decay topology always has the effect of
selecting indistinguishable 
$u\bar{u}\; u\bar{u} ,\; d\bar{d}\; d\bar{d}$ or $s\bar{s}\; s\bar{s}$
subcomponents of the states B and C, as required.

{\it With isospin symmetry:} If we assume isospin symmetry for $u,d$, then by G--parity conservation $G_{A} = G_{B}G_{C}$. Since
$G_{H} = C_H^0 (-1)^{I_H}$, we obtain $C_{A}^0 = (-1)^{I_A+I_B+I_C}$ because
the C--parities of B and C are identical. It can, however, by
explicit calculation be verified that $f=(-1)^{I_A+I_B+I_C}$ (see Appendix). Hence $f = (-1)^{I_A+I_B+I_C} 
= C_A^0$. 

{\sf Examples:} $\Pi^{\pm} \rightarrow \pi^{\pm}\pi^{0},\; \pi^{\pm}\eta,\; \pi^{\pm}\eta^{'}$.

\subsection{Connected four--quark coupling \protect\plabel{four}}

We now discuss a four--quark state A, which is 
{\it not} a molecular bound state of two mesons.
The possible four--quark decay topologies are
\figfour\ -- \figeight.
Here we focus on the connected topologies \figfour\ -- \figsix.

{\it Indistinguishable flavours:} 
{\sf Examples:} $\Omega,\; s\bar{s}s\bar{s},\; K\bar{K},\; D\bar{D},\; B\bar{B},D\; _s\bar{D}_s,\; B_s\bar{B}_s\rightarrow \eta^{'}\eta$,

\noindent where e.g. $K\bar{K}$ includes $K^+K^-$ and $K^0\bar{K}^0$.
Another application is to flavour components of state A e.g.
$u\bar{u}u\bar{u},\; d\bar{d}d\bar{d} \rightarrow \pi^{0} \eta,\; \pi^{0} \eta^{'}$.

{\it With isospin symmetry:} If we assume isospin symmetry for $u,d$, the arguments
are similar to \S \ref{hy}, noting that $f = (-1)^{I_A+I_B+I_C}$
(see Appendix) \cite{tech}.
 
{\sf Examples:} $\Pi^{\pm},\T^{\pm} \rightarrow \pi^{\pm}\pi^{0}$;\eh  
$\Pi^{\pm} \rightarrow \pi^{\pm}\eta,\; \pi^{\pm}\eta^{'}$; \eh 
\eh $\Pi^{0}\rightarrow\pi^{0}\eta,\;\pi^{0}\eta^{'}$,

\noindent where $\T$ denotes an isotensor.

{\bf Symmetrization selection rule II:} (With isospin symmetry) Connected decay and production in
topology 1 and 4 -- 6 to two
$J=0$ hybrid or conventional mesons B and C which are identical in all respects, except possibly flavour,
vanish. The processes should involve only $u,d$ quarks, and isospin symmetry
is assumed. 
A, B or C may be charged, but the hybrid or four--quark state A should have a neutral isopartner with $J^{PC} = 
0^{+-}, 1^{-+}, 2^{+-}, 3^{-+}, \ldots$ \cite{tech}.

Four--quark rules can also be applied to meson sea components, i.e.
to the connected meson sea topologies
\figfour\ -- \figsix. Assuming isospin symmetry and
states B and C that are the same in all respects except possibly flavour, $u,d$ isoscalar sea corrections to the
connected meson decay topology \figone\ vanish, as long as the
corresponding four--quark decays vanish. Non--vanishing decays arise if $s\bar{s}$ sea components
are allowed in a $u,d$ meson, or $u\bar{u},\; d\bar{d}$ sea in $s\bar{s}$, 
e.g. in the channels $\eta^{'}\eta,\; \eta^{'}\pi$ or $\eta\pi$. In this case
the dominant decay is expected from the quark rearrangement  
topology \figfive\ ($u\bar{u}s\bar{s}\rightarrow u\bar{u}\; s\bar{s}$
or $d\bar{d}s\bar{s}\rightarrow d\bar{d}\; s\bar{s}$), 
because symmetrization arguments are invalid for this topology.

\subsection{Non--connected coupling \plabel{noncon}}

We now study the non--connected
topologies \figtwo\ and \figseven.

{\it Indistinguishable flavours:} 
{\sf Examples:} In $\Pi^{0} \rightarrow \pi^{0} \eta,\; \pi^{0} \eta^{'}$;\ehh $\Eta,\; s\bar{s}\rightarrow \eta^{'}\eta$
 contributions from ${u\bar{u}}\; \\ {u\bar{u}},\;  {d\bar{d}}\; {d\bar{d}},\; 
{s\bar{s}}\; {s\bar{s}}$ components vanish. In $\Eta \rightarrow
 \eta^{'}\eta$ contributions from $\eta_{u\bar{u}+d\bar{d}}^{'}\;\eta_{u\bar{u}+d\bar{d}}$ components
vanish.

For the topologies \figthree\ and \figeight, and the glueball topology \fignine,
it na\"{\i}vely appears that
$\cabc$ is not topologically distinct from $\ca$, invalidating the application of
symmetrization arguments. Although there exists diagrams in perturbative QCD with this
property, the majority of diagrams have $\cabc$ topologically distinct from $\ca$.
For the latter diagrams we proceed to apply symmetrization arguments. 
Symmetrization selection rules where states B and C are in a ``half doughnut'' 
topology (as in topology 3b) can be shown to apply only for decays already known to vanish by CP convervation,
so we  only proceed to consider states B and C in a ``raindrop'' topology  (as in topology 3a).
From the Appendix $f=1$. Since states B and C have identical C--parities, we have $C_A =1$. So  $f = 1 = C_A$. 

{\sf Examples:} Neutral isoscalar hybrids ($\Omega,\; s\bar{s}, \; c\bar{c},\;
b\bar{b}$), four--quark states ($\Omega,\; s\bar{s}s\bar{s},\;
K\bar{K},\; D\bar{D},\; \\ B\bar{B},\;  D_s\bar{D}_s,\; B_s\bar{B}_s$) and
glueballs
$\rightarrow \eta^{'}\eta, \;
\eta_c\eta,\; \eta_c\eta^{'},\; \eta_b\eta,\; \eta_b\eta^{'},\; \eta_c\eta_b$.

{{\bf Symmetrization selection rule III:} Non--connected decay and production in topologies \figthree\ and \figeight\ -- \fignine\
of a $J^{PC} = 1^{-+}, 3^{-+}, \ldots$ hybrid, four--quark state
or glueball to
two $J=0$ hybrid or conventional mesons B and C which are identical in all respects, except possibly flavour,
vanish. The statement only holds when the $B\rl C$ exchanged diagram is topologically distinct
from the original diagram. Isospin symmetry is not assumed.

\section{Breaking of symmetrization selection rules \plabel{break}}

If we do not assume isospin symmetry, 
the possibility of different strengths of pair creation for different flavours
does not break the selection rules. 

Suppose that the states B and C have some factorizable property\footnote{
These must be interactions clearly happening within states B and C, 
and hence associated with B and C, 
distinct from the remainder of the interaction topology.} $F_H$, which
can be factored in front of the amplitude as $F_BF_C$. 
The arguments in Eq. \ref{eqw} would still be valid, since
$F_BF_C$ is invariant under $B\rl C$, even if $F_B\neq F_C$. 
Particularly, states have energy dependence
$F_H = \exp{iE_H t}$ due to time translational invariance. 
Hence different energies or masses for B and C does not
explicitly break the validity of the arguments.  

It is
clear that states B and C with different internal structure,
indirectly related to them
having different masses and energies, would break the symmetrization selection rules.
Corrections of this nature are found to be small in models 
\cite{page95light,kalashnikova94,penesize}
as long as the $J=0$ states B and C have the same radial excitation, 
and substantial otherwise.
This is accord with expectations since we expect different radial
excitations in B and C to invalidate the selection rules.
When off--shell states B and C are allowed, breaking of the rules could
be more substantial \cite{page95light}, enabling off--shell meson exchange as a 
potentially significant exotic hybrid, four--quark or glueball
production mechanism, e.g in $\pi N\rightarrow J^{PC}N$ with low energy $\pi$ exchange.

\section{Summary of symmetrization selection rules}

For connected topologies production and decay of neutral exotic $1^{-+},3^{-+}\ldots$ hybrid 
mesons and four--quark states to two $J=0$
states (hybrid or conventional mesons), e.g. pseudoscalar mesons, which are identical in all respects except possibly flavour, vanish. 
The same is true for charged and neutral
$0^{+-},1^{-+},2^{+-},3^{-+}\ldots$ states A in decays involving only $u,d$ quarks
if isospin symmetry is invoked \cite{tech}.  
For non--connected topologies, vanishing 
production and decay result for $1^{-+}, 3^{-+}, \ldots$ hybrid mesons, four--quark states
and glueballs to two $J=0$ states, which are identical in all respects, except possibly flavour. 
For topologies 2 and 7 this only applies to certain flavour components of the two $J=0$ states,
and for topologies 3 and 8 -- 9 there are conditions on the diagrams. 
All symmetrization rules are broken if the 
internal structure of B and C differs, but do not depend explicitly on the energy and mass differences
between B and C.

A special case is decays of hybrid or four--quark 
$0^{+-},1^{-+}, \ldots$\ehh $\Pi^{\pm},\T^{\pm} \rightarrow \pi^{\pm}\pi^0$ 
which vanish by isospin symmetry in {\it all} possible topologies to which it contributes
(topologies \figone, \figfour\ and \figsix), including isoscalar sea components
in the case of hybrid A \cite{tech}.

The selection rules derived in this letter go beyond well--known
selection rules, because they depend
upon the specific flavour content of the states, and on the
production or decay topology. For ground
state pseudoscalar mesons B and C, selection rules are found for 
$\eta^{'}\eta,\; \eta\pi,\;\eta^{'}\pi$ \cite{lipkin2,lipkin1}, 
$\pi^{\pm}\pi^{0},\;\eta_{c}\eta,\;\eta_{c}\eta^{'},\;\eta_{b}\eta,\;
\eta_{b}\eta^{'}$ and $\eta_{c}\eta_{b}$. 
We found three categories of
symmetrization selection rules. 
Firstly, in the absence of isospin symmetry selection rules
result when B and C have indistinguishable flavour components, e.g. $Q\bar{Q}\;
Q\bar{Q}$. 
Secondly, in the case of isospin
symmetry, selection rules are found to apply to
states B and C containing a neutral flavour--mixed hybrid or
conventional meson with flavour content $u\bar{u} + d\bar{d}$ or $u\bar{u} - d\bar{d}$. 
The selection rules result from
cancellations of amplitudes containing either the
$u\bar{u}$ component or the $d\bar{d}$ component of the
neutral flavour--mixed meson. In this way the relative sign between
the $u\bar{u}$ and $d\bar{d}$ components is sampled.
When we sum the amplitude $\ca$ and the
$B \lr C$ amplitude $\cabc$, the one amplitude picks the $u\bar{u}$
component and the other the $d\bar{d}$ component. 
Thirdly, for non--connected
``raindrop'' topologies we found in the absense of isospin symmetry that the flavour overlap is always
invariant under $B  \lr C$, leading to selection rules for
e.g. $\eta_c\eta_b$ states B and C.

\section{Comments and Phenomenology}

Assuming the same internal structure for $\eta,\; \eta^{'}$ and $\pi$, 
we predict the connected decays of valence and $u,d$ sea
components in hybrid $1^{-+}\rightarrow \eta\pi,\; \eta^{'}\pi$ to be negligible \cite{tech}. 
If the OZI suppression of non--connected decays of mesons can be extrapolated to
hybrids, a small non--connected contribution is expected. It is significant that QCD sum rule
calculations consistently predict a tiny $\eta\pi$ mode, e.g. $\sim$ 0.3 MeV (versus 600 MeV for
$\rho\pi$ and 300 MeV for $K^*K$) \cite{latorre} and small $\eta^{'}\pi$ of 4 MeV \cite{latorre} or
3 MeV (versus $\rho\pi$ of 270 MeV and $K^*K$ of 8 MeV) \cite{narison}. The relative 
size\footnote{Contributions other than topology \protect\figone\
to hybrid $1^{-+} \rightarrow \eta\pi,\; \eta^{'}\pi$ are discussed
in refs. 
\protect\cite{lipkin1,tanimoto83}.} of
$\eta\pi$ and $\eta^{'}\pi$ is consistent with a selection rule 
based on SU(3) flavour symmetry \cite{meshkov,lipkin1}. 
Ref. \cite{govaerts} also notes that $\eta\pi,\; \eta^{'}\pi$
are ``suppressed'' relative to $\rho\pi$ of $10-100$ MeV. In addition, ref. \cite{narison} notes
that $\pi\pi$ is ``suppressed'', consistent with the claim of this letter that within 
isospin symmetry, this connected decay should vanish even for sea components \cite{tech}. 

We note that if $0^{+-},1^{-+},\ldots$ light $u,d$ four--quark systems 
exist, not only are their (dominant) quark rearrangent topologies
\figfour\ and \figfive\ to pseudoscalars  suppressed, but also  
topology \figsix\ \cite{tech}. Hence decays only happen through (suppressed) non--connected topologies,
confirmed by a model calculation \cite{penesize}. 
We hence na\"{\i}vely expect e.g. the $\eta\pi$ mode of 
a $1^{-+}$ exotic be similar whether it is a hybrid or four--quark state. 
It has, however, been noted \cite{lipkin1} that $u\bar{u},\; d\bar{d}$ components of a four--quark state
can in perturbation theory be expected to mix substantially via single gluon exchange with
$s\bar{s}$, although flavour mixing of this kind has been found to be $\lapprox 10\%$
in a model calculation \cite{semay}. Presence of $s\bar{s}$ components 
would allow quark rearrangement decay via  
topology \figfive\, which is not forbidden by symmetrization rules. Measurement of 
$\eta\pi,\; \eta^{'}\pi$ decay hence samples the strength of the $s\bar{s}$ component
in a $u,d$ four--quark state. 
A four--quark state would on general grounds be expected to have a larger total width than a hybrid due to quark rearrangement topologies to non--pseudoscalars.
Thus a wide $1^{-+}$ wave could be interpreted as a four--quark state. 
If the $s\bar{s}$ component of a four--quark state is small, the state may have a typical
mesonic width\footnote{Assuming a small $\rho\pi$ coupling. Modes to $K^*K,\; \rho\eta,\;  f_2\pi, \; f_1\pi,\;  b_1\pi$ 
are near the edge of phase space.}, otherwise it is expected to be wide.
A candidate state $\hat{\rho}(1405)$
with width $180\pm 20$ MeV, possibly decaying to $\eta\pi$ but absent 
in $\rho\pi$ has been reported \cite{pdg}. There is
recent preliminary evidence \cite{bnl} for a resonance with similar width and decay patterns. 
If the $1^{-+}$ state is indeed significantly produced, 
the $\eta\pi$ mode may discriminate against the hybrid interpretation,  
since only the (suppressed) non--connected topology \figtwo\ contributes. However, the mode
may be due to a $s\bar{s}$ component in a four--quark state. Since the $s\bar{s}$ components
in $\eta$ and $\eta^{'}$ are nearly the same, and due to  P--wave phase space, we expect $\eta^{'}\pi < \eta\pi$ \cite{lipkin1}. 


A subset of the rules has explicitly been shown\footnote{
The connected decay $1^{-+}\rightarrow\eta\pi$ vanishes in quenched Euclidean QCD with   
isospin symmetry, assuming no final state interactions and a $t\rightarrow \infty$
limiting procedure which isolates only the ground state $1^{-+}$.} to arise
in QCD field theory \cite{peneqcd}. 
It would be a challenge for lattice gauge theory and Dyson--Schwinger techniques
to see if agreement is found with this result, and to estimate the size of 
non--connected topologies when unquenching the calculation.

\vspace{.5cm}


Discussions with A. Afanasiev, S.-U. Chung, F.E. Close, H.J. Lipkin, O. P\`{e}ne,
J.C. Raynal, P. Sutton and S.-F. Tuan are acknowledged.

\appendix

\section{Appendix: Flavour Overlaps}

The flavour state is 

\beqn
|H\rangle = \sum_{h\bar{h}} H_{h\bar{h}} |h\rangle |\bar{h}\rangle
\hspace{1cm} \mbox{where} \hspace{.4cm} H_{h\bar{h}} = \langle I_H I_H^z | \frac{1}{2}h \frac{1}{2}-\bar{h}\rangle
(-1)^{\frac{1}{2}-\bar{h}}
\eeqn
and $|\frac{1}{2}\rangle = u,\; |-\frac{1}{2}\rangle = d,\; |\bar{\frac{1}{2}}\rangle = \bar{u},\;
|-\bar{\frac{1}{2}}\rangle = \bar{d}$. This just yields the usual $I=1$ flavour $-u\bar{d}, \;\frac{1}{\sqrt{2}}(u\bar{u}-d\bar{d}), \; d\bar{u}$ for $I^z = 1,0,-1$ and $\frac{1}{\sqrt{2}}(u\bar{u}+d\bar{d})$ for $I=0$. The advantage of this
way of identifying flavour is that any pair creation or annihilation 
that takes place  do so with $I=0$ pairs $\frac{1}{\sqrt{2}}(u\bar{u}+d\bar{d}) = \frac{1}{\sqrt{2}}\sum_{h\bar{h}}\delta_{h\bar{h}}|h\rangle |\bar{h}\rangle$ being formed
out of the vacuum, making the operator trivial.

For the connected decay of topology \figone\ the flavour overlap $\cf$ is 

\beqn
\sum_{a\bar{a}b} A_{a\bar{a}}B_{a\bar{b}}\delta_{b\bar{b}}C_{b\bar{a}}
= \sum_{a\bar{a}b}(-1)^{\frac{1}{2}-b} \;\langle I_A I_A^z | \frac{1}{2}a \frac{1}{2}-\bar{a}\rangle
\;\langle I_B I_B^z | \frac{1}{2}a \frac{1}{2}-b\rangle
\;\langle I_C I_C^z | \frac{1}{2}b \frac{1}{2}-\bar{a}\rangle
\eeqn 
which can easily be shown under $B\rl C$ to give the sign $f=(-1)^{I_A+I_B+I_C}$. 

For four--quark states A we are free to decompose
the four quarks in two different ways
in terms of two quark--antiquark pairs. The flavour can be decomposed as

\beqna \label{f1} {\mbox{Topology }}  \figfour\ \& \;\figsix: & & 
\sum_{I_{Q\bar{q}}^zI_{q\bar{Q}}^z}
\;\langle I_A I_A^z | I_{Q\bar{q}} I_{Q\bar{q}}^z I_{q\bar{Q}} I_{q\bar{Q}}^z 
\rangle \; |A^{Q\bar{q}}\rangle |A^{q\bar{Q}}\rangle  \\ \label{f2} 
{\mbox{Topology }}  \figfive\ : & & 
\sum_{I_{Q\bar{Q}}^zI_{q\bar{q}}^z}
\;\langle I_A I_A^z | I_{Q\bar{Q}} I_{Q\bar{Q}}^z I_{q\bar{q}} I_{q\bar{q}}^z 
\rangle |A^{Q\bar{Q}}\rangle \; |A^{q\bar{q}}\rangle  
\eeqna
where we summed over all isospin projections. 
In the quark rearrangement topologies 4 and 5 it is convenient to choose a flavour decomposition 
for A which makes the overlap with B and C trivial. We obtain the flavour overlaps $\cf$

\beqna
{\mbox{Topology }} \figfour\ :  \sum_{I_{Q\bar{q}}^zI_{q\bar{Q}}^z} 
\langle I_A I_A^z | I_{Q\bar{q}} I_{Q\bar{q}}^z I_{q\bar{Q}} I_{q\bar{Q}}^z \rangle\;
\delta_{I_{Q\bar{q}}I_{B}} \delta_{I_{Q\bar{q}}^z I_{B}^z}
\delta_{I_{q\bar{Q}}I_{C}} \delta_{I_{q\bar{Q}}^z I_{C}^z}  \nonumber \\ {\mbox{Topology }} \figfive\ :  \sum_{I_{Q\bar{Q}}^zI_{q\bar{q}}^z} 
\langle I_A I_A^z | I_{Q\bar{Q}} I_{Q\bar{Q}}^z I_{q\bar{q}} I_{q\bar{q}}^z \rangle\;
\delta_{I_{Q\bar{Q}}I_{B}} \delta_{I_{Q\bar{Q}}^z I_{B}^z}
\delta_{I_{q\bar{q}}I_{C}} \delta_{I_{q\bar{q}}^z I_{C}^z}  \nonumber
\eeqna

\vspace{-0.9cm}

\beqna
\lefteqn{{\mbox{Topology }}  \figsix\ :  \sum_{I_{Q\bar{P}}^z I_{P\bar{Q}}^z} 
\langle I_A I_A^z | I_{Q\bar{P}} I_{Q\bar{P}}^z I_{P\bar{Q}} I_{P\bar{Q}}^z \rangle
\sum_{a\bar{a}b\bar{b}c\bar{c}} A^{Q\bar{P}}_{a\bar{b}}\delta_{b\bar{b}}A^{P\bar{Q}}_{b\bar{a}}
B_{a\bar{c}}\delta_{c\bar{c}}C_{c\bar{a}}\nonumber } \\ & & 
= \sum_{I_{Q\bar{P}}^zI_{P\bar{Q}}^z}
\langle I_A I_A^z | I_{Q\bar{P}} I_{Q\bar{P}}^z I_{P\bar{Q}} I_{P\bar{Q}}^z \rangle 
\sum_{a\bar{a}bc} (-1)^{1-b-c}\;
\langle I_{Q\bar{P}}I_{Q\bar{P}}^z | \frac{1}{2} a  \frac{1}{2} -b \rangle 
\nonumber\\ & &  \hspace{2.14cm}\eqntimes  
\langle I_{P\bar{Q}}I_{P\bar{Q}}^z | \frac{1}{2} b  \frac{1}{2} -\bar{a} \rangle \; 
\langle I_B I_B^z | \frac{1}{2} a  \frac{1}{2} -c \rangle\;
\langle I_C I_C^z | \frac{1}{2} c  \frac{1}{2} -\bar{a} \rangle 
\label{overl}
\eeqna
The four--quark states in Eqs. \ref{f1} and \ref{f2} are characterized by $I_A,\; I_A^z$ and the
isospins of the two quark--antiquark pairs, generically referred to as $I_X$ and $I_Y$.
In Eq. \ref{overl} denote $I_{Q\bar{q}},\; I_{Q\bar{Q}},\; I_{Q\bar{P}}$ by $I_X$ and 
$I_{q\bar{Q}},\; I_{q\bar{q}},\; I_{P\bar{Q}}$ by $I_Y$ in Eq. \ref{overl}. 
Write the four--quark state as $|I_A I_A^z I_X I_Y \rangle$. It can be seen
by explicit computation that if $I_X = I_Y$, then $f=(-1)^{I_A+I_B+I_C}$ under $B\rl C$ for each of the
expressions in Eq. \ref{overl}. 
When $I_A = 0$, the physical state is a linear combination of $|0\: 0\: 0 \: 0 \rangle$
and $|0\: 0\: 1 \: 1 \rangle$. For $I_A = 2$, the physical state is $|2\: I_A^z \: 1 \: 1 \rangle$.
So in both cases $I_X = I_Y$. When $I_A=1$, the physical state is a linear combination of 
$|1\: I_A^z 1\: 1 \rangle , \; |1\: I_A^z 1\: 0 \rangle$ and $|1\: I_A^z 0\: 1 \rangle$.
For each of the expressions in Eq. \ref{overl} it can be shown that 
$|1\: I_A^z 1\: 0 \rangle \rightarrow (-1)^{I_A+I_B+I_C}\; |1\: I_A^z 0\: 1 \rangle$ under $B\rl C$. Defining new
states $|\pm\rangle \equiv |1\: I_A^z 1\: 0 \rangle \pm |1\: I_A^z 0\: 1 \rangle$, we see that
under $B\rl C$, $|\pm\rangle \rightarrow \pm (-1)^{I_A+I_B+I_C}|\pm\rangle$. 
The above statements about the behaviour of $\cf$ under $B\rl C$ are true in any decomposition of A,
hence also in the ``diquonium'' decomposition, where pairs of two quarks and two antiquarks are used. 
The advantage of this decomposition is that neutral states $|\pm\rangle$
would be eigenfunctions of charge conjugation \cite{semay}, as required. Hence, when $I_A=1$, a
physical state is either the ``positive'' linear combination of 
$|1\: I_A^z 1\: 1 \rangle$ and $|+\rangle$, or the ``negative'' linear combination of 
$|1\: I_A^z 1\: 1 \rangle$ and $|-\rangle$. The former gives $f=(-1)^{I_A+I_B+I_C}$ under $B\rl C$,
and the latter has no proper symmetry under $B\rl C$. In summary, for $I_A = 0$ or 2, or
for ``positive'' $I_A = 1$ states, $f=(-1)^{I_A+I_B+I_C}$ under $B\rl C$.
 
For the ``raindrop'' configurations in topologies \figthree\ and \figeight\ -- \fignine\  
the part of the flavour overlap $\cf$ containing reference to B and C is 

\beqn
\sum_{a\bar{a}} A_{a\bar{a}}\delta_{a\bar{a}}\sum_{b\bar{b}} B_{b\bar{b}}\delta_{b\bar{b}}
= Tr(A)\; Tr(B)
\eeqn
which under $B\rl C$ gives $f=1$.

\end{document}